\documentclass[epj-spec]{svjour}

\usepackage{graphics}
\usepackage{color} 
\usepackage[normalem]{ulem}
\usepackage{colortbl}

\newcommand{\be}{\begin{equation}}
\newcommand{\ee}{\end{equation}}
\newcommand{\bea}{\begin{eqnarray}}
\newcommand{\eea}{\end{eqnarray}}

\begin{document}

\title{Phase transition in the scalar noise model
of collective motion in three dimensions}

\author{Bal\'azs G\"onci\inst{1}\fnmsep\thanks{\email{gonci@angel.elte.hu}}
\and M\'at\'e Nagy\inst{1} \and Tam\'as Vicsek\inst{1,2}}

\institute{
Department of Biological Physics, E\"otv\"os University,
P\'azm\'any P.\ stny.\ 1A, H-1117 Budapest, Hungary.
\and
Biological Physics Research Group of HAS,
P\'azm\'any P.\ stny.\ 1A, H-1117 Budapest, Hungary
}

\abstract{
We consider disorder-order phase transitions in the three-dimensional version of the
scalar noise model (SNM) of flocking. Our results are analogous to those found
for the two-dimensional case \cite{nagy} and \cite{vicsek}.
For small velocity ($v\leq 0.1$) a continuous, second-order phase transition
is observable, with the diffusion of nearby particles being isotropic.
By increasing the particle velocities the phase transition changes to first
order, and the
diffusion becomes anisotropic. The first-order transition in the latter case
is probably caused by the interplay between anisotropic diffusion and periodic
 boundary
conditions, leading to a boundary condition dependent symmetry
breaking of the solutions.
}
\maketitle
\section{Introduction} 
\label{intro}
The collective motion of animals is a fascinating phenomenon, resulting
from the occasionally very simple interactions between individuals.
Self-propelled particle (SPP) models
\cite{vicsek,czirok,csahok,tonertu1,tonertu2,tonertu3,czirok_3d,gregoire}
play a crucial role in understanding of the dynamics of these
biological systems \cite{animal},
be them bird flocks \cite{birds}, fish schools \cite{fish}, insect
swarms \cite{insect} or bacteria aggregates \cite{bacteria}. A common
feature of these groups is the onset of collective motion without a
leader.

The two dimensional scalar noise model (SNM) is able to reproduce some
fundamental properties of these collective motions.
In spite of its simplicity,
the SNM reproduces the emergence of cooperative motion from a disordered
phase in the absence a "leader" particle \cite{vicsek}.
The velocity of particles $v$ provides a control parameter which
switches between SPP behaviour ($v>0$) and equilibrium type models ($v=0$).
It means that there is a kinetic phase transition between a disordered state
and an ordered state; in the latter one most of the particles move
nearly in the same direction.

The original model of Ref. \cite{vicsek} assumes a constant
velocity $v$ for each particle. The particles in every time step
adopt the average velocity of particles within a neighbourhood
radius $R$. Besides, there is a random noise, an angle added to
the direction of the velocity in order to involve realistic noise
elements. However, the dynamic properties of this system, being
subject to random velocity fluctuations, are likely to depend
strongly on the dimensionality of the space, as it was discussed
in Ref. \cite{czirok_3d}. In \cite{czirok_3d} the transition to
the ordered phase was shown to be analogous to the one obtained
for 2d. However, in addition, a major difference was found between
the SPP and the equilibrium models: in the static ($v=0$) case the
system does not order for densities below a critical value close
to 1 (which corresponds to the percolation threshold of randomly
distributed spheres) while in the SSP ordering is found for all
velocities. In this paper we extend the investigation of the phase
transitions in the SNM model to three dimensions (3dSNM), and
investigate the phase transition at different particle velocities.

We mainly use the definitions of the original SNM.
We assume a three-dimensional space with a linear size of $L$
with periodic boundary conditions. Both the interaction radius
and the time step are set to unity ($R=1$, $\Delta t=1$), with the velocity
and the linear size measured in units equal to R.
The number of particles $N$ is also fixed, hence the density of particles
is defined as $\rho = N/L^3$. Initially the particles are randomly distributed
within a cube, with constant $v$ absolute velocities in random directions
chosen uniformly distributed in a sphere.
Particle velocities are updated simultaneously at each time step.
A given particle assumes the average direction of motion of the particles in
its local neighbourhood with some uncertainty, thus the divergence between
the direction of its velocity and the direction of the local average
velocity is an angle $\xi$. The noise tag $\xi$ is a random value in the
interval $\eta[-\pi,\pi ]$, chosen by assuming a uniform probability distribution,
where $\eta$ is the strength of the noise.
The velocities of the particles are updated obeying the rule
\be
{\bf v_i}(t+\Delta t) = v \cdot {\cal M}({\bf e},\xi) \cdot {\bf N}(\langle {\bf v}(t)\rangle_{R,i}),
\ee
where $\langle {\bf v}(t)\rangle_{R,i}$ is the average velocity vector
within the interaction radius $R$ of particle $i$, including particle $i$ itself,
${\bf N}({\bf u}) = {\bf u}/ \vert {\bf u} \vert$, ${\cal M}({\bf u},\theta)$ is
rotation tensor representing a rotation about a unit vector ${\bf u}$ for angle
$\theta$. ${\bf e}$ is a random unit vector chosen uniformly distributed
perpendicular to ${\bf N}(\langle {\bf v}(t)\rangle_{R,i})$,
thus ${\bf e}$ defines a random axis for the rotation.
The magnitude of the velocity of the particles
is fixed to $v$.
The positions of particles were updated according to
\be
{\bf x_i}(t+\Delta t)= {\bf x_i}(t)+{\bf v_i}(t)\Delta t.
\ee
In order to describe the phase transition and to characterise the collective
behaviour of the particles, we define the average normalised velocity as
\be
\varphi= \frac{1}{Nv}\vert \sum_{i=1}^N{\bf v_i}\vert ,
\ee
which is our order parameter. $\varphi$ is zero
when the individual directions of
particles velocity are chosen randomly (for infinite systems),
and is equal to one if all particles move in the same direction.

\section{Kinetic Phase Transition} 
\label{sec2}
We have investigated the statistical behaviour of the 3dSNM model using the
probability distribution function (PDF) of the order parameter $\varphi$,
and the Binder cumulant $G$.
The Binder cumulant \cite{binder}, defined as $G=1-<\varphi^4>/3<\varphi^2>^2$,
measures the fluctuations of the order parameter. Therefore, the function of
$G$ versus $\eta$ has different shapes. At the occurrence of a first-order phase
transition, $G$ has a definite minimum, while in case of a second-order phase
transition the characteristic minimum does not occur.

We have investigated the phase transition for $L=64, N=32768,
\rho=1/8$ with velocities $v=0.03, 0.1, 0.5, 1.0, 3.0, 10.0$. For
the small velocity regime ($v=0.03, 0.1$) we have found that the
PDF of the order parameter has one definite peak, while the Binder
cumulant does not exhibit a characteristic minimum (Fig. 1/a. and
Fig. 2/a.). The particles aggregate into clusters of different
sizes as shown in two dimensions \cite{huepe}, a typical snapshot
of these simulations is shown on Fig. 3/a. As we increase the
velocity to a higher regime ($v=0.5, 1.0, 3.0$) the shape of both
the PDF and the Binder cumulant versus $\eta$ function change. The
PDF of the order parameter has two peaks near the critical $\eta$
value, and the Binder cumulant has a definite minimum, albeit it
has large deviations (Fig. 1/b. and Fig. 2/b.). Typical snapshots
of these simulations are shown on Fig. 3/b. and Fig. 3/c. In
summary, by increasing the particle velocities the phase
transition changes from second-order to first-order similarly to
what was described in Ref. \cite{aldana}. Most of our results have
been obtained for a fixed value of the density. For a few trial
runs with other (higher) densities we found analogous behaviours.
The main difference is that the critical noise changes as a
function of the density \cite{czirok_3d}.

At the velocity value $v=10.0$ we find another change. The PDF charts
are similar to those observed at the small velocity regime. The Binder
cumulant does not exhibit a characteristic minimum, but it shows
a sharper change between the ordered and disordered phases (Fig. 4.).
We assume that the model`s behaviour changes to a continuous mean field model
behaviour at this velocity.
The particle velocity compared to $L$ is large enough to carry the
information in a few steps everywhere in the simulation field. Thus every particle in
the system has information about the entire field, therefore the
system is likely to behave as a continuous mean field model.
\begin{figure}[t!]
\centerline{\resizebox{0.9\textwidth}{!}{\rotatebox{0}{\includegraphics{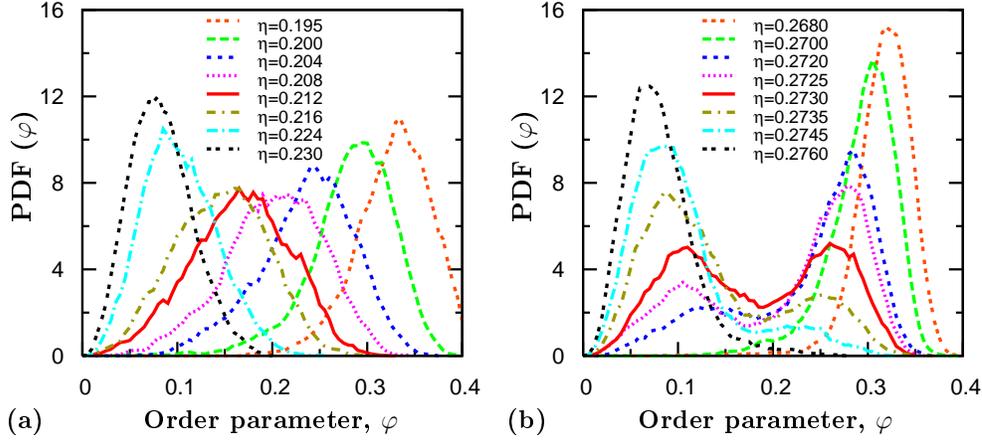}}}}
\caption{ Probability distribution function (PDF) of the order parameter
$\varphi$ for noise values $\eta$ around the critical point. A curve with
one single peak indicates a {\it second-order phase transition}.
Two peaks indicate a
{\it first-order phase transition}. (a) Continuous, second-order phase
transition for $v=0.1$. (Analogous behaviour can be observed at $v=0.03$.)
(b) First-order phase transition at $v=0.5$. (The same behaviour can
be observed for $v=1.0$ or $v=3.0$.) Other parameters: $\rho=1/8$, $L=64$,
$N=32768$, $T=10^6$.}
\label{fig:Pdf_v0.1_0.5}
\end{figure}
\begin{figure}[t!]
\centerline{\resizebox{0.9\textwidth}{!}{\rotatebox{0}{\includegraphics{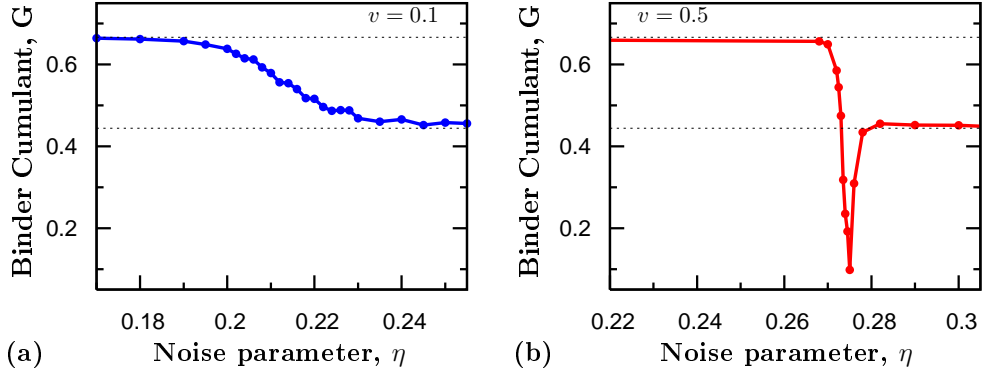}}}}
\caption{ The Binder cumulant $G$ as a function of the noise $\eta$.
The absence of a minimum point in $G$ (smooth behaviour) indicates a
{\it second-order phase transition}, whereas its presence $G$ indicates
a {\it first-order phase transition}. Values of $G$ in the ordered phase
and in the disordered phase are denoted by horizontal lines at $G=2/3$
and $G=4/9$, respectively.
(a) The behaviour of the Binder cumulant confirms that the phase
transition at $v=0.1$ is of
second-order. (The same behaviour can be observed for $v=0.03$.)
(b) The definite minimum around $\eta=0.275$ means a first-order phase
transition at $v=0.5$. (The same behaviour can be observed for $v=1.0$ or
$v=3.0$)
Other parameters: $\rho=1/8$, $L=64$, $N=32768$, $T=10^6$.}
\label{fig:Binder_v0.1_0.5}
\end{figure}
\begin{figure}[t!]
\centerline{\resizebox{0.9\textwidth}{!}{\rotatebox{0}{\includegraphics{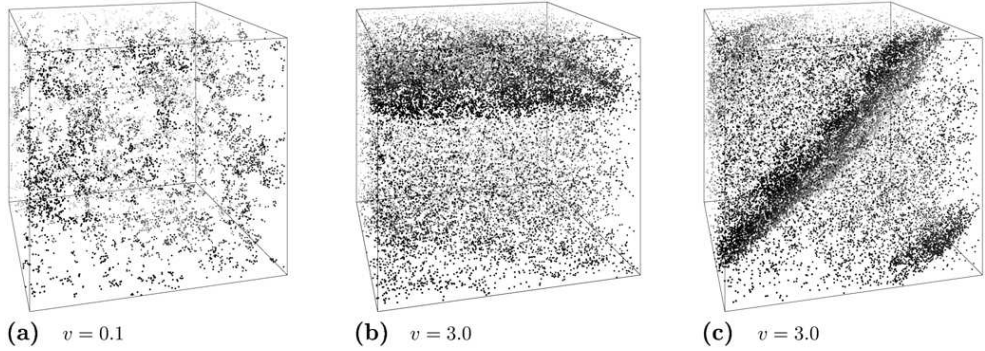}}}}
\caption{Snapshots of the spatial distribution of particles in the 3dSNM.
The colour shading shows the distance of particles, particles in the far
fade towards light gray.
(a) A typical snapshot of the system in the small velocity regime
($v=0.1$). The isolated flocks moving coherently with a characteristic
steady state size coexisting with a background gas of particles.
(b) Particles move parallel to the edges of the simulation cube in the large
velocity regime ($v=3.0$), and a density wave emerges.
(c)  Density wave moves diagonally ($v=3.0$).
Other parameters: $\rho=1/8$, $L=64$, $N=32768$, $\eta=0.1$.}
\label{fig:snapshot}
\end{figure}
\begin{figure}[t!]
\centerline{\resizebox{0.5\textwidth}{!}{\rotatebox{0}{\includegraphics{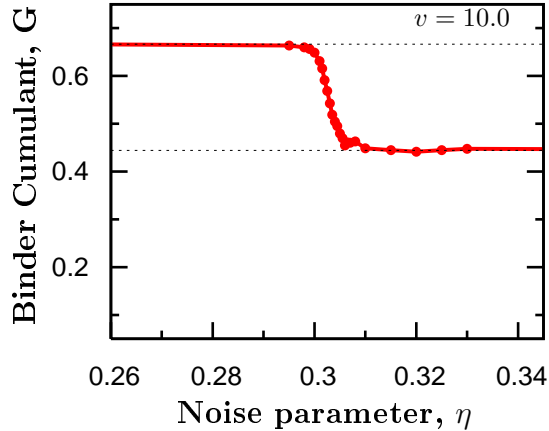}}}}
\caption{ The Binder cumulant $G$ as a function of the noise at particle
velocity $v=10.0$.
$G$ has a smooth behaviour, indicating a {\it second-order phase transition},
but the value decreases significantly faster at the transition point.
Values of $G$ in the ordered phase and in the disordered phase are
denoted by horizontal lines at $G=2/3$ and $G=4/9$, respectively.
Other parameters: $\rho=1/8$, $L=64$, $N=32768$, $T=10^6$.}
\label{fig:Binder_v10.0}
\end{figure}
\section{Particle diffusion}
\label{sec3} In order to investigate the 3dSNM in more details and
interpret the unusual change of the order of transition as a
function of velocity, we have assessed the temporal divergence of
neighbouring particles \cite{nagy}. We locate pairs in the entire
system, and in every simulation step we compute the average
distance of the pairs \be r(t) = \frac{1}{N_p} \vert
\sum_{i=1}^{N_p} {d_i(t)}\vert , \ee where $r_i(t)$ is the
distance between members of the $i$th pair at time $t$, and $N_p$
is the number of pairs with $r_i(t_0)<R$ initially at time $t_0$.
We evaluated both the parallel and the perpendicular components of
the divergence, where parallel means a direction along the average
velocity vector of all particles. To eliminate artifacts of the
periodic boundary condition, we only use the data where
$r(t)\leq\frac{1}{12}L$. Typical averaged square displacement
curves as a function of time can be seen in (Fig. 5.).
\begin{figure}[t!]
\centerline{\resizebox{0.9\textwidth}{!}{\rotatebox{0}{\includegraphics{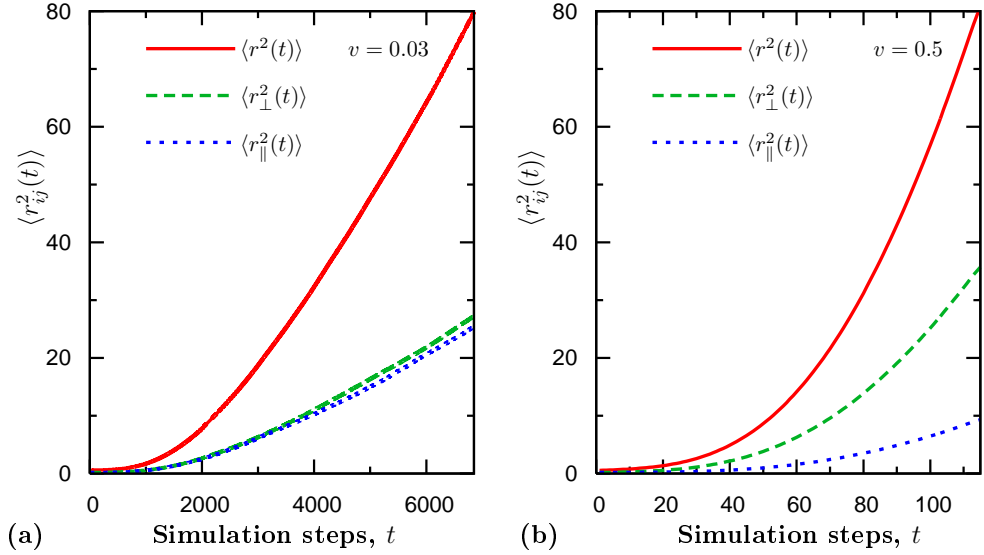}}}}
\caption{Averaged relative square displacements of neighbouring
particles as a function of time. The perpendicular square displacement
is the sum of two perpendicular directions. In order to
compare this with the parallel (to $\langle v_i \rangle_i)$)
one, we divide the square displacement
by two. (a) The diffusion is isotropic, $v=0.03$. At $v=0.1$ we can observe
a little anisotropy in the diffusion, hence we name $v=0.1$ the border
velocity between the large and small velocity regimes.
(b) The diffusion is anisotropic; it is significantly
larger in the perpendicular direction ($\langle r^2_\bot \rangle$) than
in the parallel direction ($\langle r^2_\Vert\rangle$).
On log-log scale we can observe the super-diffusion behaviour, as written
in the text (graphs not shown).
Other parameters: $\rho=1/8$, $L=64$, $N=32768$, $T=5*10^5$.}
\label{pairDist2}
\end{figure}

A super-diffusive \cite{gregoiretu1} behaviour occurs at medium
diffusion times in the ordered state in both small and large
velocity regimes, where $\alpha$ is greater than one ($\langle r^2
\rangle(t) \sim t^\alpha$). At low noise levels three different
mean square displacements can be observed. At short diffusion
times particles keep their relative positions to each other; the
diffusion is almost frozen ($\alpha \sim 0$). We assume that on
this time scale most particles stay in the same locally ordered
flock where they were when the pairs were formed. At intermediate
times, when particles have time to change their flocks, a
super-diffusive behaviour occurs. At this time scale, initially
neighbouring particles diverge from each other linearly, due to
the effects of the different flocks they belong to. In other
words, the mean spatial separation of the particles increases
approximately linearly or faster (e.g., turning away) with time
(the flocks perform random walk only on a larger time scale),
leading to a super-diffusive behaviour with an exponent of $\alpha
\geq 2$. Finally, on large time scale the flocks themselves
exhibit random walk movement and the distance of the pairs
increases with an exponent $\alpha = 1$.

In the large velocity regime we have also found
an anisotropy in diffusion while
in the small velocity regime the diffusion is isotropic. The
diffusion in the perpendicular direction is greater than in the parallel one
(shown in Fig. 5.). The ratio of the square displacement components
$r^2_\bot/r^2_\Vert$ shows the level of anisotropy; it increases
by the particle velocity. Because of the greater diffusion value
in the perpendicular direction at large velocities
density waves occur in three dimensions, similarly to the two dimensional,
original SNM model \cite{nagy}. The density waves
in three dimensions are planes, the direction of the normal vector is
parallel to the average velocity of all particles.
It means that although we have found
a definite minimum in the Binder cumulant, and the
PDF graphs have two peaks, we can not exclude the possibility that these
results are only
an artificial outcome of the large velocity regime, caused by the periodic
boundary conditions. The self-appending density waves may induce an
artificial strengthening in
the ordered state, causing the first-order phase transition-like
behaviour ultimately.
\section{Symmetry breaking in the directional distribution}
\label{sec4}
The symmetry breaking in the diffusion in the large velocity regime is
discussed in the previous section; it can cause the emergence
of density waves.
The density waves may have other signs too. If they occur, the directional
distribution must have significant symmetry breaking, because moving
in the direction parallel to the edges of the cube corresponds to a
more stable state.
With smaller possibility diagonal density waves also occur.

We have investigated the probability distribution function (PDF)
of the direction of the averaged velocity of the particles, and we
have found uniform distribution in the small velocity regime and
anisotropic distribution at the large velocity regime (Fig. 6.).
These results support our assumption that density waves do not
occur at small particle velocities, until density waves emerge for
larger particle velocities.
\begin{figure}[t!]
\centerline{\resizebox{0.9\textwidth}{!}{\rotatebox{0}{\includegraphics{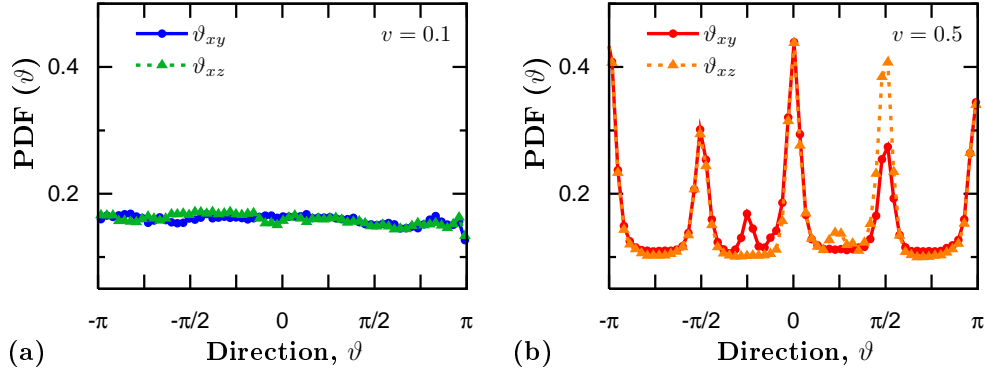}}}}
\caption{ The directional distribution of the average velocity.
$\vartheta_{xy}$, $\vartheta_{xz}$
denote the direction of the average velocity projected to plane $xy$, $xz$
respectively. (a) At low
particle velocities the distribution is uniform ($v=0.1$).
(b) At large particle velocities symmetry breaking occurs ($v=0.5$).
The peaks at $\vartheta=k*\frac{\pi}{2}$ (where $k$ is integer) indicate that
the particles movement is most likely
along the edges of the cube. One can observe
smaller peaks at $\vartheta=\frac{\pi}{4}$. Corresponding density waves move
diagonally and have higher linear size, thus they less likely than the
parallel ones.
Other parameters: $\rho=1/8$, $L=64$, $N=32768$.
}
\label{PhiDirDistr}
\end{figure}

We emphasize that the original purpose of the Vicsek et. al.
\cite{vicsek} model is to interpret the collective movement of
migrating cells, flocking birds or other biological systems, where
the self-propelled particle velocity is small compared to the
ratio of their interaction radius and reaction time $v\ll R/\tau$.
With the parameters used this means that the velocity must be
small compared to unity. Our investigation showed that velocity
with the parameters used in three dimensions is $v\leq 0.1$.
\section{Conclusions}
\label{conclusion}
We have found that the behaviour of the 3dSNM
is similar to that of the SNM in two dimensions
in many aspects. In the small velocity regime
the order-disorder transition is of second-order, continuous phase transition
just as in two-dimensions \cite{vicsek,czirok}. At this velocity values
artificial symmetry breaking did not occur, thus the results have physical
relevance to the biological systems.

In the large velocity regime a further discontinuous symmetry
breaking occurs in the direction of the average velocity, not
observed before for the 3d case. This indicates the emergence of
density waves \cite{gregoire} (showed earlier in two dimensions
\cite{nagy}), caused by the anisotropic diffusion of initially
neighbouring particles, due to the periodic boundary conditions.
Hence we can not draw any conclusions about the physical behaviour
of the 3dSNM in the large velocity regime. Finite size effects in
the case of SPP models are very delicate, since they are due to
the fact that in the ordered regime the particles run through the
cell in a time linearly proportional to L, so they "feel" the two
opposite side of the system in a relatively short time interval.
Going to larger sizes does not solve this problem, because,
although the above "scanning" time somewhat increases, the
relative magnitude of the inherent fluctuations decreases and the
effect of the rectangular symmetry of the simulational cell shows
up even more markedly. This is a paradoxical situation needing
further exploration. Finally, if the velocity and the linear size
of the system are of the same order of magnitude, the system
exhibits a continuous mean field like behaviour.

\begin{acknowledgement}
We would like to thank P\'eter Szab\'o for the helpful remarks on the
manuscript.
This work has been supported by the Hungarian Science Foundation
(OTKA), grant No. T049674 and EU FP6 Grant "Starflag".
\end{acknowledgement}

\end{document}